# A non-reflecting metamaterial slab under the finite-embedded coordinate transformation

**Il-Min Lee[1], Seung-Yeol Lee[1], Kyoung-Youm Kim[2], and Byoungho Lee[1*]**

[1]*National Creative Research Center for Active Plasmonics Applications Systems*
*Inter-University Semiconductor Research Center and School of Electrical Engineering*
*Seoul National University,Gwanak-Gu Gwanakro 599, Seoul 151-744, Korea*
[2]*Department of Optical Engineering, Sejong University, Gunja-Dong, Gwangjin-Gu, Seoul 143-741, Korea*

*[*] byoungho@snu.ac.kr*

**Abstract:** From the explicit solutions of Maxwell's equations under the coordinate transformation, the conditions for non-reflecting boundaries for the two-dimensionally propagating light waves, in a finite-embedded coordinate transformation metamaterial slab are derived in cases of extended two-dimensional. By exploring several examples, including some reported in the literatures and some novel developed in this study, we show that our approach can be used to efficiently determine the condition in which a finite-embedded coordinate transformed metamaterial slab is non-reflecting.

©2010 Optical Society of America

## 1. Introduction

Coordinate transformation optics (hereafter referred to as transformation optics) is a recently introduced concept for tailoring electromagnetic waves to achieve certain desirable characteristics via the use of specially designed metamaterials [1]. The technique is based on the form-invariance of Maxwell's equations under smooth and differentiable coordinate transformations [2]. The mathematically defined transformations are cast into corresponding changes of constitutive material parameters and consequently, of electromagnetic entities, such as electromagnetic fields, currents, and charge distributions. The invisibility cloak has been proposed [1] as an innovative example of this, and many related studied both experimental [3] and theoretical, have been proposed. In addition to cloaking devices, including the carpet cloak [4], acoustic [5] and matter wave [6] cloaks and an anti-cloak [7], other interesting concepts of coordinate transformation optics such as adaptive beam bands and beam expanders [8], field concentrators [9], source transformation devices [10], and illusion optical devices [11] have been reported as well.

From a topological view point, transformation optics can be categorized into two types. In the first type of transformation, the spatial domain of the transformation is ideally localized within a finite space: in other words, the interaction of light waves with the transformed region is designed to be seen as if there is a localized object such as free-space (invisibility cloak), ground mirror (carpet cloak) or other object (illusion optics). Therefore, the properties of outgoing light waves from such transformed media, at least in ideal cases, recover those of their original states. In contrast, another type of transformation optical device has been proposed, i.e. finite-embedded coordinate transformation optical devices [12]. In this type of transformation, the metrics of space are not necessarily recovered into those of the original space. In many cases, the main objective of transformed optical media is to modify the outgoing wave fronts. The beam shifter and divider [12], and beam expander [8] are examples of this.

The concept of finite-embedded coordinate transformation optics has stimulated some interest in the reflection conditions at the interfaces between transformed and untransformed media. Due to some slight vagueness in the reflectionless condition between these media suggested in the original introductory paper [12], the issue of the necessary and/or sufficient conditions for finite-embedded media to be transparent has been discussed [8, 13-17]. The non-reflecting condition derived in Ref. [13] covers three-dimensionally transformed media with non-planar boundaries. The derived condition given in Eq. (11) in Ref. [13] states that if the boundary coordinate in the transformed medium can be expressed as the rotation of the original space with a constant displacement vector added, then the boundary is non-reflecting. However, in this report we will show that a non-reflecting coordinate transformed boundary can exist for a skewed-and-expanded transformation (see Section 3.2 and Fig. 4) which cannot be covered by Eq. (11) in Ref. [13]. Another general study on the boundary conditions for generalized transformed media was also reported [14]. Although our study is restricted to more specific cases of extended two-dimensional problems, we provide more explicit conditions and illustrative examples throughout this paper which were not addressed in Ref. [14]. Recently, studies on the mapping method that can make the two-dimensional transformation optical slab transparent in a general way have been reported [15-17]. However, the discussions concerning this are limited only to cases of two-dimensionally transformed media and cannot deal with extended two-dimensional cases which are treated in this paper (see Section 4 and Fig. 5).

In this work, the non-reflecting boundary conditions of a finite-embedded coordinate transformed metamaterial slab will be explored. For the sake of simplicity, we imposed two restrictions in our study. The first one is that the mapping function of the coordinate transformation is not $z$-dependent, except for $z'$, which can be $x$- and $y$-dependent. This is more general than the two-dimensional transformations treated in Refs. [15-17]. As will be seen in this study, this is not a straightforward extension of the two-dimensional case and care must be taken in deriving the non-reflecting condition. We also assumed that the incident field is defined in two-dimensional conventions, i.e., we excluded conical incidence. With this assumption, the dispersion relation in the transformed media can be given in a very simple form. The governing equation which makes transformed media transparent will be presented, along with details of the derivation procedure. Several examples, including known mapping functions in several literatures as well as some novel ones, will be discussed with numerical implementations. A discussion of the design of some novel transformed optical devices are also presented.

**2. Derivation of the governing equation**

In this section, we start with the basics of the coordinate transformation in a Cartesian coordinate system and derive an equation which describes the non-reflecting boundaries in two-dimensional finite-embedded coordinate transformation media (CTM). Let us consider a coordinate transformation in a Cartesian coordinate system, which maps the points in the original space (or the virtual space) $(x, y, z)$ to those in the physical space (or the transformed space) $(x', y', z')$ as follows:

$$(x', y', z') = (F(x, y, z), G(x, y, z), H(x, y, z)), \tag{1}$$

where $F$, $G$, and $H$ are the mapping functions of the coordinate transformation. We further assume that the mapping functions in Eq. (1) can have an inverse transformation.

Now let us consider the case where this CTM is finitely embedded as in Fig. 1. We assume that the material parameters in the original space can be different from those of the surrounding media in the physical space. As depicted in Fig. 1, the transformed medium (region *II*) fills a finite-extent of the physical space and the material parameters of both sides of the transformed medium (region *I* and *III*) can also be different.

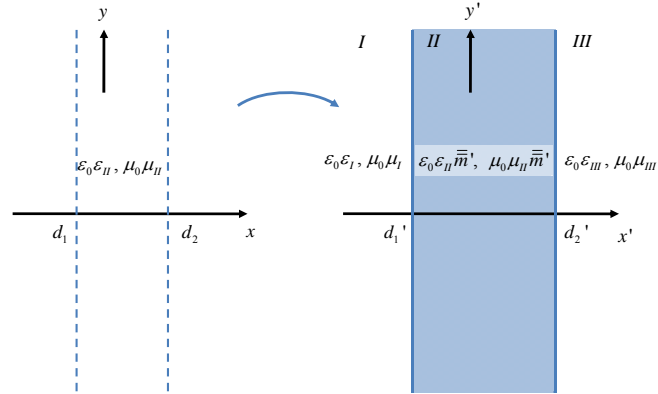

Fig. 1. Geometrical representation of a finite-embedded coordinate transformation.

For the sake of simplicity, we further assume that the mapping functions in Eq. (1) are defined in a sort of extended two-dimensional convention such that

$$F(x, y, z) = F(x, y), \ \ G(x, y, z) = G(x, y), \ \ H(x, y, z) = h(x, y)z. \tag{2}$$

This means that the mapping functions are not *z*-dependent except for $z'$ which can be *x*- and *y*-dependent by the scale factor $h(x, y)$ : this is the reason why we refer to this mapping extended two-dimensional.

From the form-invariance of Maxwell's equations, the material tensors in virtual space $\bar{\bar{m}}$ can be re-expressed in the physical space as $\bar{\bar{m}}'$ with the following relation [1]:

$$\bar{\bar{m}}^{i'j'} = A_i^{i'} A_j^{j'} \bar{\bar{m}}^{ij} / \det(A_i^{i'}), \tag{3}$$

where $\bar{\bar{m}}$ and $\bar{\bar{m}}'$ denote either the permittivity or the permeability tensor and $A_i^{i'}$ is the element of the transformation Jacobian matrix defined as

$$A_i^{i'} = \partial x^{i'} / \partial x^i. \quad (4)$$

Through this paper, we assume that the material in the virtual space is isotropic and homogeneous so that the material parameters in the original space can be expressed as $\bar{\bar{\varepsilon}} = \varepsilon_0 \varepsilon_r$ and $\bar{\bar{\mu}} = \mu_0 \mu_r$. Under this condition, the material tensor in transformed space can be defined as $\bar{\bar{\varepsilon}} = \varepsilon_0 \varepsilon_r \bar{\bar{m}}'$ and $\bar{\bar{\mu}} = \mu_0 \mu_r \bar{\bar{m}}'$ with $\bar{\bar{m}}' = AA^t / \det(A)$ where $A$ is the Jacobian matrix and the superscript $t$ denotes the matrix transpose.

Now suppose that the material tensor $\bar{\bar{m}}'$ can be expressed as

$$\bar{\bar{m}}' = \begin{bmatrix} a & b & c \\ b & d & e \\ c & e & f \end{bmatrix} \quad (5)$$

and its matrix inverse $(\bar{\bar{m}}')^{-1}$ is given as

$$(\bar{\bar{m}}')^{-1} = \frac{1}{\Delta} \begin{bmatrix} (df - e^2) & -(bf - ce) & (be - cd) \\ -(bf - ce) & (af - c^2) & -(ae - bc) \\ (be - cd) & -(ae - bc) & (ad - b^2) \end{bmatrix} = \frac{1}{\Delta} [\Delta_{ji}], \quad (6)$$

where $\Delta_{ij}$ is the matrix cofactor and $\Delta$ is the determinant of $\bar{\bar{m}}'$ given as

$$\Delta = adf - ae^2 - b^2 f - c^2 d + 2bce \quad (7)$$

from the coordinate transformation results shown in Eq. (1). Substituting Eq. (5) into Maxwell's equations with time dependency of $\exp(-j\omega t)$ gives the following wave equation in a matrix form:

$$\frac{k_0^2}{\Delta^2} \mathbf{SE} + k_0^2 n^2 \mathbf{E} = 0, \quad (8)$$

where $\mathbf{E}$ is a column vector of $(E_x, E_y, E_z)$, $n$ is the refractive index of the original material, $n_{II}^2 = \varepsilon_{II} \mu_{II}$, and $k_0$ is the free-space wavenumber. The system matrix $\mathbf{S}$ in Eq. (8) is given as

$$\mathbf{S} = \begin{bmatrix} \Delta_{12}\kappa_z - \Delta_{13}\kappa_y & -\Delta_{11}\kappa_z + \Delta_{13}\kappa_{x,II} & \Delta_{11}\kappa_y - \Delta_{12}\kappa_{x,II} \\ \Delta_{22}\kappa_z - \Delta_{23}\kappa_y & -\Delta_{12}\kappa_z + \Delta_{23}\kappa_{x,II} & \Delta_{12}\kappa_y - \Delta_{22}\kappa_{x,II} \\ \Delta_{23}\kappa_z - \Delta_{33}\kappa_y & -\Delta_{13}\kappa_z + \Delta_{33}\kappa_{x,II} & \Delta_{13}\kappa_y - \Delta_{23}\kappa_{x,II} \end{bmatrix}^2, \quad (9)$$

where $\kappa_i = k_i / k_0$ and we drop the prime notations in subscripts for simplicity. From the condition that Eq. (8) has a non-trivial solution, the determinant of $(\mathbf{S}/\Delta^2 + n^2 \mathbf{I})$, where $\mathbf{I}$ is the identity matrix, should be zero. Therefore, we arrive at the following dispersion relation for this transformed media

$$n_{II}^2 (-\Delta n_{II}^2 + a\kappa_{x,II}^2 + d\kappa_y^2 + f\kappa_z^2 + 2b\kappa_{x,II}\kappa_y + 2c\kappa_z\kappa_{x,II} + 2e\kappa_y\kappa_z)^2 / \Delta^2 = 0. \quad (10)$$

For simplicity, we assume that the incident plane wave arising from region $I$ has $k_z = 0$. Under this assumption and for a non-zero $n$, the dispersion relation in Eq. (10) can be expressed with cofactor elements as:

$$\Delta_{33}(\Delta_{22}\kappa_{x,II} - \Delta_{12}\kappa_y)^2 = \Delta_{22} n_{II}^2 \Delta^2 + \Delta_{22}(\Delta_{23}\kappa_{x,II} - \Delta_{13}\kappa_y)^2 - \Delta_{33}(\Delta_{11}\Delta_{22} - \Delta_{12}^2)\kappa_y^2. \quad (11)$$

In general, the transformed medium in region $II$ is inhomogeneous and anisotropic and thus, cannot be approximated efficiently using simple models such as a stair-case approximation. Therefore, we restrict our analysis to those cases in which the variations in the material parameters in the transformed space are smooth and differentiable. Furthermore, since our interest is to attempt to determine conditions that make this finite slab of CTM reflectionless, multiple reflections can be ignored. As a consequence, we can consider the reflections at each interface ($x'=d_1'$ and $x'=d_2'$) separately as if they are single interfaces. In addition, when we obtain the reflectionless condition at the first boundary ($x'=d_1'$), the condition required for reflectance of the second boundary ($x'=d_2'$) to be zero can be obtained very easily, i.e., we can adopt the condition obtained at the first boundary with an appropriate change in material parameters from *I* to *III*.

From an analysis of eigenvalues and eigenvectors of this transformed media under the assumption of a plane monochromatic wave, it is apparent that the TE- and TM-polarized fields are eigen-bases of this transformed material. Therefore, we can decompose the problem into TE- and TM-polarized cases. Now let us consider the reflectance of the TM-polarized plane monochromatic wave [ $\mathbf{E}=(E_x,E_y,0)$ and $\mathbf{H}=(0,0,H_z)$ ] incident on the first boundary. The magnetic field components in the immediate vicinity of the first boundary are readily expressed as

$$H_{z,I} = \exp[j(k_{x,I}x+k_y y)] + r\exp[j(-k_{x,I}x+k_y y)], \quad (12)$$

$$H_{z,II} = t\exp[j(k_{x,II}x+k_y y)], \quad (13)$$

where *r* and *t* are complex coefficients of the reflectance and the transmittance, respectively. From Maxwell's equations, the tangential electric field in region *I* can be easily obtained as

$$E_{y,I} = \frac{k_{x,I}}{k_0 \varepsilon_I}\{\exp[j(k_{x,I}x+k_y y)] - r\exp[j(-k_{x,I}x+k_y y)]\}. \quad (14)$$

The tangential electric field in region *II* can be obtained similarly with the material parameters given in Eq. (5) as

$$E_{y,II} = \frac{\Delta_{22}k_{x,II} - \Delta_{12}k_y}{k_0 \varepsilon_{II}\Delta} t\exp[j(k_{x,II}x+k_y y)]. \quad (15)$$

From the continuity of the tangential components of the electromagnetic field at the boundary, we have

$$r = \frac{\Delta\varepsilon_{II}\kappa_{x,I} - \varepsilon_I(\Delta_{22}\kappa_{x,II} - \Delta_{12}\kappa_y)}{\Delta\varepsilon_{II}\kappa_{x,I} + \varepsilon_I(\Delta_{22}\kappa_{x,II} - \Delta_{12}\kappa_y)}, \quad (16)$$

$$t = \frac{2\Delta\varepsilon_{II}\kappa_{x,I}}{\Delta\varepsilon_{II}\kappa_{x,I} + \varepsilon_I(\Delta_{22}\kappa_{x,II} - \Delta_{12}\kappa_y)}. \quad (17)$$

Therefore, to make *r* zero, we have

$$\Delta\varepsilon_{II}\kappa_{x,I} = \varepsilon_I(\Delta_{22}\kappa_{x,II} - \Delta_{12}\kappa_y). \quad (18)$$

To make the medium reflectionless for all incidence angles (i.e. for arbitrary values of $k_y$), Eq. (18) should be satisfied for the arbitrary values of $k_y$. Considering Eqs. (11) and (18), we find that such conditions can be fulfilled only if the following condition is satisfied by the coordinate transformation (see the Appendix):

$$(A_x^{x'}A_y^{z'} - A_y^{x'}A_x^{z'})(A_x^{x'}A_x^{z'} + A_y^{x'}A_y^{z'}) = 0. \quad (19)$$

When Eq. (19) is satisfied, the condition for which the extended two-dimensional finite embedded coordinate medium is non-reflecting can be obtained as (see the Appendix):

$$\Delta_{22}^2\mu_{II}^2 = \Delta_{33}\mu_I^2, \quad (20)$$

which can be expressed in Jacobian elements as

$$(A_z^{z'})^2[(A_x^{x'})^2 + (A_y^{x'})^2]^2 \mu_{II}^2 = (A_x^{x'}A_y^{y'} - A_y^{x'}A_x^{y'})^4 \mu_I^2, \quad (21)$$

and further simplified in an alternate form as

$$(A_z^{z'})[(A_x^{x'})^2 + (A_y^{x'})^2] = \pm\rho(A_x^{x'}A_y^{y'} - A_y^{x'}A_x^{y'})^2, \quad (22)$$

where $\rho = \mu_I / \mu_{II}$.

The condition for the second interface can be readily obtained by changing $\mu_I$ in Eq. (20) with $\mu_{III}$:

$$\Delta_{22}^2 \mu_{II}^2 = \Delta_{33}^2 \mu_{III}^2. \tag{23}$$

Let us consider the case $\mu_I^2 \neq \mu_{III}^2$. Even for this case, we can still generally say that the boundaries can be made non-reflecting when the mapping function at each boundary satisfies the conditions in Eq. (20) or (23) separately and the mapping functions in between change smoothly. Under this condition, and from the duality for the TE-polarized case, we have the following form for the condition for the non-reflecting boundary:

$$(A_z^{z'})[(A_x^{x'})^2 + (A_y^{x'})^2]\big|_i = \pm \rho_i (A_x^{x'} A_y^{y'} - A_y^{x'} A_x^{y'})^2 \big|_i. \tag{24}$$

where $\rho_i = \varepsilon_i / \varepsilon_{II}$ for TE- and $\rho_i = \mu_i / \mu_{II}$ for TM-mode light waves and $i$ denotes either *I* or *III* and we used the notation $\big|_i$ to distinguish left and right boundaries.

In fact, the solution sets that satisfy Eq. (20) are not identical to those satisfying Eq. (18): in squaring Eq. (18), another set of solutions is introduced. In fact, this latter set of solutions makes the denominator of Eq. (16) or (17) zero, which physically means that these solutions are those for the excitation of the waveguide mode propagating along the *y*-direction with a propagation constant $k_y$ satisfying the squared form of Eq. (18). In practical situations, the excited waveguide modes under the condition of Eq. (21) are surface modes (formed only when the two media (*I* and *II*) have opposite signs for permittivity or permeability) which can be excited by the incidence of evanescent waves. Therefore, if we exclude this specific case of surface wave excitation that requires both opposite signs and the incidence of evanescent waves, the conditions in Eq. (24) can be regarded as necessary and sufficient conditions for a non-reflecting boundary. Otherwise, these conditions become necessary ones to make the finite-embedded CTM non-reflecting.

In Eq. (24), the $\pm$ sign means that the signs of the material parameters between the transformed and the background media can be freely selectable. However, as stated above, choosing opposite signs can result in the excitation of the surface waves at the interface, and if it is not inevitable, selecting the same sign as the surrounding media (i.e., choosing + in Eq. (24)) is recommended.

Equation (24) can be used as a guideline for the design of a planar finite-embedded CTM interface. Since the differential equation given in Eq. (24) is quite general, providing a general solution is difficult. Instead, we will explore several simple test solutions and show the utility and validity of the derived conditions in the next sections.

## 3. Two-dimensional examples

The simplest example satisfying Eq. (19) is the case where the transformation is totally two-dimensional:

$$A_x^{z'} = A_y^{z'} = 0. \tag{25}$$

For this case, we can assume that the mapping functions given in Eq. (1) can be expressed in the following forms:

$$x'(x,y) = A(x) + B(y) + F(x)G(y) + c_1, \tag{26}$$

$$y'(x,y) = C(x) + D(y) + H(x)K(y) + c_2, \tag{27}$$

$$z'(x,y) = c_3 z, \tag{28}$$

where $c_i$ ($i = 1, 2, 3$) are constants.

By substituting Eqs. (26)-(28) into (24) we have

$$c_3[(A_x + F_x G)^2 + (B_y + FG_y)^2]\big|_i = \pm \rho_i[(A_x + F_x G)(D_y + HK_y) - (B_y + FG_y)(C_x + H_x K)]^2 \big|_i, \tag{29}$$

where $|_i$ denotes the left and right boundaries and the subscript $\eta$ means the partial derivative for the variable $\eta$, i.e., $X_\eta = \partial X(x,y)/\partial \eta$ for $\eta = x$ or $y$. In following sub-sections, we consider several examples for the mapping functions given in Eqs. (26)-(28).

*3.1. Beam shifter and beam expander*

The beam shifter and expander based on coordinate transformation optics are shown schematically in Fig. 2.

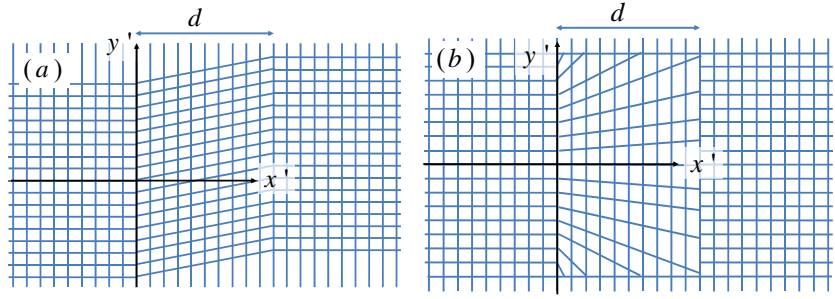

Fig. 2. Geometrical representation of the beam shifter and expander.

The mapping function of the beam shifter in its simple form can be found in Ref. [12] and is given as follows:

$$x' = x, \quad y' = y + \tan\theta x, \quad z' = c_3 z, \tag{30}$$

where the upper and lower limits in the transformed region are ignored, for the simplicity of the discussion.

By comparing the conventions given in Eqs. (26)-(28), we can set

$$A(x) = x, \quad F(x) = 1, \quad C(x) = x\tan\theta, \quad D(y) = y, \quad K(y) = 1, \tag{31}$$

with other functions not specified in Eq. (31) so as to all be zero. Substituting these in Eq. (29) gives

$$c_3 = \pm \rho_i. \tag{32}$$

For the case of $c_3 = 1$, this result supports the widely-known fact that if the impedance of the medium in the virtual space for the given polarization is matched to those of the background media in the physical space, the shift transformation causes no reflection at the boundary [12].

Let us turn our attention to the beam expander. The simplest form of the mapping function for the beam expander can be given as [8]

$$x' = x, \quad y' = m(x)y, \quad z' = c_3 z, \tag{33}$$

where the magnification coefficient $m(x)$ is assumed to have values of 1 and a constant value $M$ at the first ($x = 0$) and second ($x = d$) boundaries, respectively. In a simple implementation, the $x$-dependent magnification coefficient can be defined in a linear function of $x$ as follows:

$$m(x) = (M-1)(x-d)/d + M, \tag{34}$$

where $d$ is the $x$-directional thickness of the transformed region.

From Eqs. (26)-(28), we can set:

$$A(x) = x, \quad F(x) = 1, \quad H(x) = m(x), \quad K(y) = y, \tag{35}$$

with other functions that are not specified as zero. Substituting these in Eq. (29) gives,

$$c_3 = \pm \rho_i [m(x_i)]^2. \tag{36}$$

From the definition that $c_3$ is a constant, the above result implies that, if the square of the magnification factor $m(x)$ at the boundaries becomes different from $\pm c_3/\rho_i$, the coordinate expansion in the transversal direction (along $y$-direction) will cause reflections. Therefore, to make the interfaces non-reflecting, regions *I* and *III* should be filled with material that can satisfy Eq. (36).

*3.2. Rotated and skewed-and-expanded transformation*

Now let us consider another example in which the mapping functions in Eqs. (26)-(28) have the form of the combined characteristics of the longitudinal and transversal expansions. We will examine a simple mapping function given as

$$x' = \alpha x + \beta y, \quad y' = \chi x + \delta y, \quad z' = z, \tag{37}$$

where $\alpha$, $\beta$, $\chi$, and $\delta$ are all non-zero constants. From the conventions in Eqs. (26)-(28), we have the following:

$$A(x) = \alpha x, \quad B(y) = \beta y, \quad C(x) = \chi x, \quad D(y) = \delta y, \tag{38}$$

with the other functions and constants not specified as zero. From the condition of Eq. (29) we have

$$(\alpha^2 + \beta^2) = (\alpha\delta - \beta\chi)^2, \tag{39}$$

where we assumed that the impedances between media *I*, *II*, and *III* are matched and $\pm\rho$ is dropped for the sake of simplicity.

Initially, consider the case in which $\delta = \alpha$ and $\chi = -\beta$. For this case, Eq. (39) can be expressed as

$$(\alpha^2 + \beta^2) = (\alpha^2 + \beta^2)^2. \tag{40}$$

Therefore, from the assumption of non-zero constants, we have $(\alpha^2 + \beta^2) = 1$. This condition can be generally satisfied by setting $\alpha = \cos\theta$ and $\beta = \sin\theta$. Hence the mapping in Eq. (37) can be expressed as

$$x' = \cos\theta x + \sin\theta y, \quad y' = -\sin\theta x + \cos\theta y, \quad z' = z, \tag{41}$$

which corresponds to the two-dimensional rotation of the coordinates. Actually, the rotation of the coordinates does not change the material parameter in Eq. (5) because $\bar{\bar{m}}'$ for this transformation is a unit diagonal.

However, Eq. (39) can be satisfied in various other ways than $\delta = \alpha$ and $\chi = -\beta$. For example, by choosing the values of the constants in Eq. (39) as $\alpha = 3$, $\beta = 4$, $\chi = 1$, and $\delta = 3$, this gives the mapping of

$$x' = 3x + 4y, \quad y' = x + 3y, \quad z' = z, \tag{42}$$

and a metric tensor of

$$\bar{\bar{m}}' = \begin{bmatrix} 5 & 3 & 0 \\ 3 & 2 & 0 \\ 0 & 0 & 0.2 \end{bmatrix}, \tag{43}$$

which satisfies Eq. (39).

This coordinate transformation given in Eq. (42) maps the geometrical grids of the virtual space in the Cartesian coordinates into a skewed and expanded one as presented in Fig. 3. In Fig. 3, the dotted and circled lines correspond to lines that are parallel to the *x*-axis and *y*-axis before the transformation, respectively, and the thin grid lines denote the grid lines in the $x'$-$y'$ coordinates.

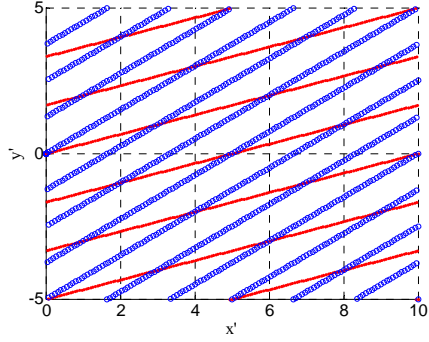

Fig. 3. Transformed geometrical grid made by the skewed and expanded transformation given by Eq. (42).

In Fig. 4, we present the electric field distributions obtained from a commercial software program based on the finite element method [18]. In this figure, the field radiating from a point (line) source propagates through a finite coordinate transformed slab which satisfies the condition Eq. (39) by the mapping function in Eq. (42) without reflections. In this calculation, the thickness of the coordinate transformed slab is $4\lambda$, where $\lambda$ is the free-space wavelength.

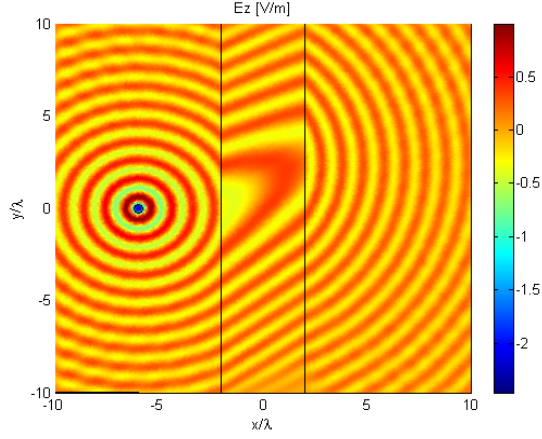

Fig. 4. Field distributions for the skewed-and-expanded coordinate transformation.

*3.2. Further examples*

Some interesting discussions on the reflectionless condition of the finite-embedded CTM has been reported [17]. In that study, the following mapping function was considered:

$$x' = au(x), \quad y' = y/u_x(x) + v(x), \quad z' = z, \tag{44}$$

which can be expressed using the conventions in Eqs. (26)-(28) as

$$A(x) = au(x), \quad C(x) = v(x), \quad H(x) = 1/u_x(x), \quad K(y) = y, \tag{45}$$

in which the other functions and constants are all zero. In this case, using the newly-derived condition in Eq. (29), we can obtain the following condition for a non-reflecting boundary:

$$(au_x)^2 = (au_x/u_x)^2, \tag{46}$$

which corresponds to

$$u_x = \pm 1, \tag{47}$$

and this coincides well with the results reported in Ref. [17].

Let us consider another example similar to Eq. (44): the following form of transformation

$$x' = au_x(x), \quad y' = y/u(x) + v(x), \quad z' = z, \tag{48}$$

which can be expressed using the conventions in Eqs. (26)-(28) as

$$A(x) = au_x(x), \quad C(x) = v(x), \quad H(x) = 1/u(x), \quad K(y) = y, \tag{49}$$

with other functions and constants as zero. Using the condition in Eq. (29), we have the condition of the non-reflection boundary as

$$(au_{xx})^2 = (au_{xx}/u)^2, \tag{50}$$

which corresponds to

$$u = \pm 1 \tag{51}$$

at the boundaries. From the mapping functions in Eqs. (44) and (48), it can be found that mapping functions which enable transparency can have various other forms than that introduced in Ref. [12].

### 4. Extended example: an extended two-dimensional case

*4.1. Reflectionless beam expander*

Now let us consider more complicated cases which satisfy Eq. (19). What we will examine is the case in which the mapping functions are given as

$$x' = x, \quad y' = m(x)y, \quad z' = c(x)z. \tag{52}$$

This satisfies Eq. (19) for $A_y^{z'} = A_y^{x'} = 0$. The mapping functions in Eq. (52) correspond to the case in which the scaling of the axes is applied simultaneously in two tangential directions. For this case, the material tensor $\bar{\bar{m}}'$ is given as

$$\bar{\bar{m}}' = \frac{1}{cm} \begin{bmatrix} 1 & m_x y & c_x z \\ m_x y & (m_x y)^2 + m^2 & m_x y c_x z \\ c_x z & m_x y c_x z & (c_x z)^2 + c^2 \end{bmatrix}. \tag{53}$$

If we assume that $k_z = 0$, Eq. (22) is applicable to this case, we arrive at the following condition for the non-reflecting boundaries:

$$c(x) = \pm \rho_i [m(x)]^2. \tag{54}$$

The electric field distributions and the corresponding values of the material parameters are shown in Fig. 5.

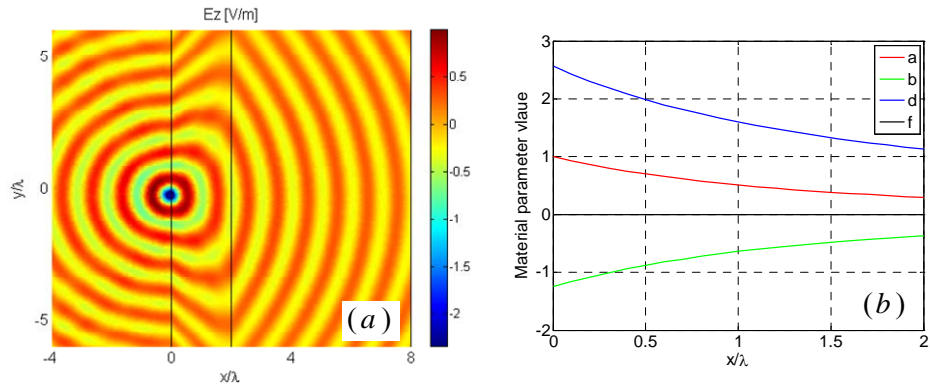

Fig. 5. Numerical results for the example of the two-dimensional beam expander without any reflection. The normalized field distribution at $z = 0$ plane is depicted in (a). The results in (b) show the material parameters ($\bar{\bar{m}}$) along the line $y = 5\lambda$.

The results presented in Eq. (54) and in Fig. 5 are contradictory to conventional knowledge on coordinate transformations based a beam expander. Until now, it was generally believed that a non-reflecting beam expander cannot be produced except for the case where the indices of the surrounding media are selected so as to match the square value of the magnification factor $m(x)$ at the boundaries (i.e. see the related discussions in Ref. [8]). However, as evidenced by Eq. (54), this is not true if we restrict the magnification to be applied in only one of the transversal directions (along the direction of *y*-axis in this case) and ignore the conical incidences ($k_{z,I} = 0$).

In Fig. 5, we assumed that all the untransformed media are free-space ($\rho = 1$) and the *x*-dependent magnification factor is a linear function given in Eq. (34) with the thickness of the slab *d* of 2λ and the magnification factor at the second interface *M* as 1.5. The calculated field distribution shows that the slab for the beam expander expands the beam, with almost no reflections, at the interfaces. The small observable ripples in the field distribution at the left-side of the CMT slab are due to numerical errors and the non-optimized nature of the gradient along the *x*-direction in the scale function $m(x)$. If the material parameters of the transformed region in the *x*-direction are carefully designed to implement adiabatic variation over a sufficiently long width, the ripples can be decreased. However, the results clearly show that the proposed method can be readily applied to the design of a reflectionless CTM.

## 5. Conclusions

A general method for creating a non-reflecting finite-embedded coordinate transformed slab, when the mapping function is defined in extended two-dimensional conventions under the condition of $k_z = 0$ and a governing equation that can be used as the determinant equation or as a guideline for the design is proposed. To verify our approach and demonstrate the validity of the strategy, we explored several examples and discussed the topological meanings of these. We hope that our approach and the discussions contained in this report will be helpful in designing or understanding finite-embedded coordinate transformed media. We also hope that the approach introduced in this study can serve as a guide or a starting point for investigations of more general cases such as conical incidences or full three-dimensional mapping functions.

**Acknowledgment**

The authors wish to acknowledge the support of the National Research Foundation and the Ministry of Education, Science and Technology of Korea through the Creative Research Initiative Program (Active Plasmonics Application Systems).

**Appendix**

In this section, we show that Eq. (19) is a necessary condition for a slab of finite-embedded coordinate transformation to be non-reflecting under the constraint of the extended two-dimensional mapping functions given in Eq. (2).

Let us start from the condition which makes the reflectance *r* zero in Eq. (18):

$$\Delta \varepsilon_{II} \kappa_{x,I} = \varepsilon_I (\Delta_{22} \kappa_{x,II} - \Delta_{12} \kappa_y). \quad (A1)$$

To make the interface non-reflecting, Eq. (A1) should be held for arbitrary value of $\kappa_y$. For this objective, we square both sides of Eq. (A1) and eliminate all $\kappa_{x,I}$ and $\kappa_{x,II}$ dependencies by using the dispersion relations for media *I* and *II*. The dispersion relation for medium *I* is expressed as

$$\kappa_{x,I}^2 = n_I^2 - \kappa_y^2, \quad (A2)$$

and that for medium *II* is expressed as in Eq. (11), which can be re-expressed as

$$[(\Delta_{22}\Delta_{33} - \Delta_{23}^2)\kappa_{x,II} + (\Delta_{23}\Delta_{13} - \Delta_{12}\Delta_{33})\kappa_y]^2 = \Delta^2[(\Delta_{22}\Delta_{33} - \Delta_{23}^2)n_{II}^2 - \Delta_{33}\kappa_y^2]. \quad (A3)$$

For any real-valued coordinate transformation $a = [(A_x^{x'})^2 + (A_y^{x'})^2 + (A_z^{x'})^2] / \det(A)$ cannot be zero. Therefore, we have $\Delta_{22}\Delta_{33} - \Delta_{23}^2 = a\Delta \neq 0$. If we assume that the coordinate transformation can be complex-valued, and hence admit those cases of $\Delta_{22}\Delta_{33} - \Delta_{23}^2 = 0$, Eq. (A3) can be expressed as

$$[(\Delta_{23}\Delta_{13} - \Delta_{12}\Delta_{33})^2 + \Delta^2\Delta_{33}]\kappa_y^2 = 0. \tag{A4}$$

Let us examine Eq. (A4) further. When $(\Delta_{23}\Delta_{13} - \Delta_{12}\Delta_{33})^2 + \Delta^2\Delta_{33} = 0$, $\kappa_y$ cannot be determined and otherwise, $\kappa_y = 0$. All these cases are contrary to the claim that Eq. (A1) holds regardless of $\kappa_y$. Therefore, we discard the case where $\Delta_{22}\Delta_{33} - \Delta_{23}^2$ is zero. Therefore, we can express $\kappa_{x,II}$ as a function of $\kappa_y$:

$$\kappa_{x,II} = \frac{1}{\Delta_{22}\Delta_{33} - \Delta_{23}^2}\left\{-(\Delta_{23}\Delta_{13} - \Delta_{12}\Delta_{33})\kappa_y \pm \Delta\left[(\Delta_{22}\Delta_{33} - \Delta_{23}^2)n_{II}^2 - \Delta_{33}\kappa_y^2\right]^{1/2}\right\}. \tag{A5}$$

When we substitute Eqs. (A2) and (A5) into Eq. (A1) and, after performing some arithmetic manipulations, we have

$$(\Delta_{22}\Delta_{33} - \Delta_{23}^2)^2\Delta^2\varepsilon_{II}^2(n_I^2 - \kappa_y^2)$$
$$= \varepsilon_I^2\left\{\Delta_{23}(\Delta_{12}\Delta_{23} - \Delta_{22}\Delta_{13})\kappa_y \pm \Delta\Delta_{22}[(\Delta_{22}\Delta_{33} - \Delta_{23}^2)n_{II}^2 - \Delta_{33}\kappa_y^2]^{1/2}\right\}^2. \tag{A6}$$

For the sake of simplicity, we cast a simplified from of Eq. (A6) by adopting proper substitutions in the coefficients:

$$A - B\kappa_y^2 = \left[C\kappa_y \pm (D + E\kappa_y^2)^{1/2}\right]^2, \tag{A7}$$

which can be expressed as

$$A - D - (B + C^2 - E)\kappa_y^2 = \pm 2C\kappa_y(D + E\kappa_y^2)^{1/2}. \tag{A8}$$

We square both sides of Eq. (A8) to obtain

$$(A - D)^2 - 2[(A - D)(B + C^2 + E) + 2C^2D]\kappa_y^2 + [(B + C^2 + E)^2 - 4C^2E]\kappa_y^4 = 0. \tag{A9}$$

This should be held for arbitrary values of $\kappa_y$. Therefore, we have the following set of equations that should be satisfied simultaneously:

$$A - D = 0, \tag{A10}$$
$$CD = 0, \tag{A11}$$
$$(B + C^2 + E)^2 - 4C^2E = 0. \tag{A12}$$

Before solving the above coupled equations, let us first examine Eq. (A11). From this equation, we can find that either $C$ or $D$ should be zero. By comparing Eqs. (A6) and (A7), we can find that Eq. (A11) can be re-expressed in a more explicit form as

$$\Delta_{23}(\Delta_{12}\Delta_{23} - \Delta_{22}\Delta_{13}) = 0 \text{ or } \Delta_{22}^2(\Delta_{22}\Delta_{33} - \Delta_{23}^2) = 0, \tag{A13}$$

where the two equations correspond to the conditions that $C = 0$ or $D = 0$, in the order of appearance.

Let us consider the latter case first. As we examined before, $\Delta_{22}\Delta_{33} - \Delta_{23}^2$ is not zero. Then if $\Delta_{22} = 0$, all $\kappa_{x,II}$ dependent terms in the dispersion relation of Eq. (A3) vanish, i.e., the value of $\kappa_{x,II}$ is undetermined. Therefore we can also discard this case. Therefore, to satisfy Eq. (A11), we find that $C = 0$. With this result, we further find from Eqs. (A10) and (A12) that $A = D$ and $B = -E$.

By comparing Eqs. (A6) and (A7) once again and using the results of $A = D$ and $B = -E$, we have following conditions:

$$(\Delta_{22}\Delta_{33} - \Delta_{23}^2)\varepsilon_{II}^2 n_I^2 = \Delta_{22}^2\varepsilon_I^2 n_{II}^2, \tag{A14}$$
$$(\Delta_{22}\Delta_{33} - \Delta_{23}^2)^2\varepsilon_{II}^2 = \Delta_{22}^2\Delta_{33}\varepsilon_I^2. \tag{A15}$$

Equations (A14) and (A15) can be further simplified by combining them as

$$\Delta_{33}n_I^2 = (\Delta_{22}\Delta_{33} - \Delta_{23}^2)n_{II}^2. \tag{A16}$$

As a final step, we substitute Eq. (A16) into Eq. (A14) to obtain the form of
$$\Delta_{33}\mu_I^2 = \Delta_{22}^2\mu_{II}^2. \tag{A17}$$

Now, let us consider the condition of $C = 0$ in more detail. As we examined in Eq. (A13), this corresponds to the case $\Delta_{23}(\Delta_{12}\Delta_{23} - \Delta_{22}\Delta_{13}) = 0$. This condition can be expressed using the Jacobian elements of the coordinate transformation:
$$(A_x^{x'}A_y^{z'} - A_y^{x'}A_x^{z'})(A_x^{x'}A_x^{z'} + A_y^{x'}A_y^{z'}) = 0. \tag{A18}$$

With Eq. (A18), we derive Eq. (19). Equation (A18) is a necessary condition for the boundary of an extended two-dimensional finite embedded coordinate medium to be non-reflecting. As a result, Eqs. (A17) and (A18) are necessary and sufficient conditions for an extended two-dimensional finite embedded coordinate medium to be non-reflecting.